\documentstyle[amssymb,aps,multicol,graphicx,prl,floats]{revtex}

\begin{document}
\draft

\twocolumn[\hsize\textwidth\columnwidth\hsize\csname @twocolumnfalse\endcsname
\author{M. E. Gershenson and Yu. B. Khavin}
\address{Serin Physics Laboratory, Rutgers University, Piscataway, NJ 08854-8019}
\author{D. Reuter, P. Schafmeister, and A. D. Wieck}
\address{Lehrstuhl fuer Angewandte Festkoerperphysik, Ruhr-Universitaet Bochum,
Universitaetsstrasse 150 \\
D-44780 Bochum, Germany}
\date{\today }
\title{Electron-Assisted Hopping in Two Dimensions}
\maketitle

\begin{abstract}
We have studied the non-ohmic effects in the conductivity of a two-dimensional system
which undergoes the crossover from weak to strong localization with decreasing electron 
concentration. When the electrons are removed from equilibrium 
with phonons, the hopping conductivity depends only on the electron temperature. 
This indicates that the hopping transport in a system with a large localization
length is assisted by electron-electron interactions rather than by the phonons. 
\end{abstract}


]

Low$-$dimensional conductors demonstrate the crossover from weak localization
(WL) to strong localization (SL) with decreasing electron concentration
and/or increasing disorder. The WL regime, where electron motion is
diffusive, and the localization and interaction effects reveal themselves as
quantum corrections to the conductivity, is well-understood now (for an
experimental review, see \cite{aags}). In the SL regime, electron transport
is due to activated hopping between the localized states \cite{es}. The
mechanism of this hopping is still under debate. On one hand, it is common
wisdom to treat this hopping as phonon-assisted (see, e.g., \cite{es,ebert,dahm}).
On the other hand, recent observations suggest that hopping is
assisted by some other mechanism rather than phonons \cite{khondaker}. 
The idea of hopping due to electron-electron scattering in conductors
with a large localization length
has been also discussed in several theoretical papers (see, e.g. 
\cite{alein-shkl,shah}).

In this Letter, we study the non-ohmic effects in the conductivity of a
two-dimensional (2d) electron gas on both sides of the WL-SL crossover.
It is shown that in either regime, diffusive or hopping, the nonlinear effects are
well described by the model of ''hot'' electrons, and  electron transport 
is controlled by the {\it electron} temperature rather than the phonon one.
This indicates that hopping transport on the
''insulating'' side of the crossover  is governed by the electron-electron
interactions.

The resistance $R$ of a two-dimensional (2d) $Si$ $\delta $-doped $GaAs$
structure has been studied as a function of the magnetic field $B$\ and the
bias current $I$ in the temperature range $T=0.05-1K$. A single $\delta $%
-doped layer with concentration of $Si$ donors $1.3$x$10^{12}$ $cm^{-2}$ is 
$0.1\mu m$ under the surface of the MBE-grown undoped $GaAs$. Two identical devices were
formed by photolithography and wet etching on a chip with dimensions $5$x$6$x%
$0.5$ $mm^{3};$ the area of each device between the voltage leads was $0.36$x%
$0.12$ $mm^{2}$. The electron concentration can be tuned by applying the
voltage $V_{g}$ to the gate electrode on top of each device. The
measurements were done in the four-probe configuration using a low-frequency ($f=7Hz$) 
lock-in technique in the range $R<10^{6}\Omega $, and a dc current
source and an electrometer for $10^{6}\Omega <R<10^{9}\Omega $.

We have observed the WL-SL crossover with reducing the electron
concentration (see the inset in Fig.3); the detailed analysis of the crossover will be given
elsewhere \cite{gkrsw}. The ''zero-bias'' dependences $R(T)$ are shown on
both sides of the crossover in Figs.1 and 2. The logarithmic dependences $%
R(T)$, observed in the WL regime, are caused by the weak
localization and interaction effects at $B=0$, and by interaction only at $%
B=0.1T$ (this field is sufficiently strong to suppress the $T$-dependence
of the WL\ correction)  \cite{aags}. Because of insufficient
filtering of the external noise, $R(T)$ is saturated 
at $T\leqslant 0.15K$. This indicates that the electrons were never cooled
below $0.1K$ in the WL regime.  
\begin{figure}[th]
\begin{center}
\includegraphics[height=3.5in, width=3in]{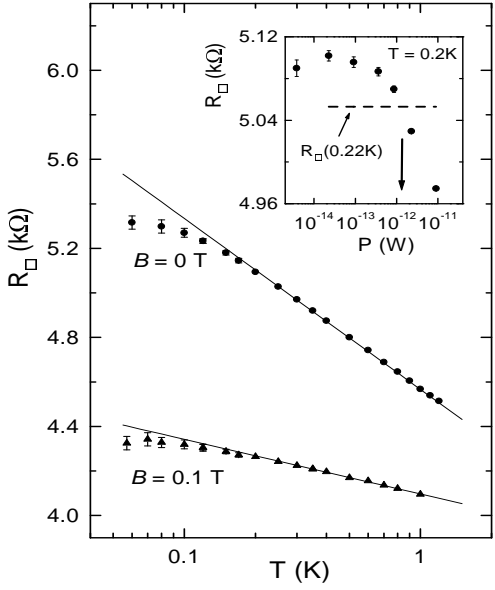}
\caption{The temperature dependences of the zero-bias $R_{\Box }$ 
in the ''metallic'' regime ($V_{g}=0.4V$, the electron
concentration $n=1$x$10^{12}cm^{-2}$) at $B=0$ and $B=0.1T$. The solid
lines are the logarithmic fit $R_{\Box }(T)=R_{\Box }(1K)\left[ 1-\protect%
\alpha \frac{e^{2}R_{\Box }}{2\protect\pi ^{2}\hbar }\ln \left( T/1K\right) %
\right] $ with $\protect\alpha =1.29$ ($B=0$) and $\protect\alpha =0.52$ ($%
B=0.1T$). The inset shows $R=V/I$ at a fixed MC temperature $T=0.2K$ and $B=0
$ versus the power $P=V\cdot I$ released in the device. The horizontal
dashed line corresponds to the zero-bias $R_{\Box }$ at $T=0.22K$.}
\label{Fig.1}
\end{center}
\end{figure}
Electron transport becomes activated on the ''insulating'' side of the WL-SL crossover.
The temperature dependence of the sheet resistance $R_{\Box }$ is usually fitted in 
this regime as 
\begin{equation}
R_{\Box }(T)=R_{0\Box }(T/1K)^{m}\exp (T_{0}/T)^{\alpha }\text{ \ ,}
\label{expfit}
\end{equation}
several combinations of $m=0-1$ and $\alpha =1/3-1$ have been reported for
different 2d systems \cite
{ebert,dahm,khondaker,hsu,goldman}. A good fit for our data at $V_{g}=-0.7V$
is provided by Eq.(\ref {expfit}) with $m=0$ and $\alpha =0.7$ (see Fig.2). Notice that in the
SL regime, where the resistance at low $T$ is by $10^{5}$ larger than $R$ in
the WL regime, saturation of $R(T)$ is absent \cite{saturation}.  
\begin{figure}[th]
\begin{center}
\includegraphics[height=3.5in, width=3in]{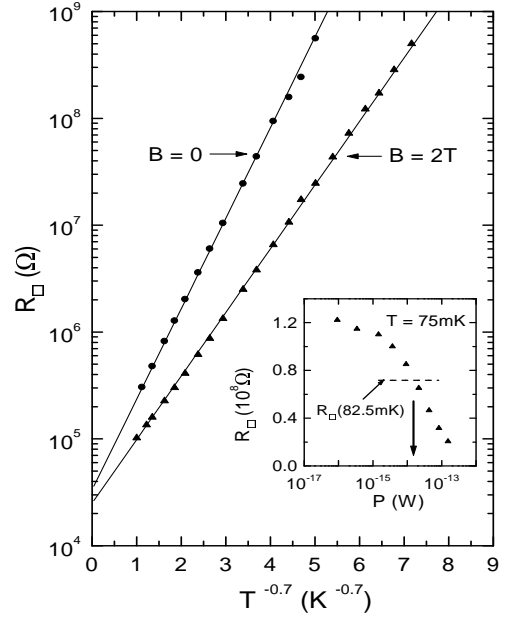}
\caption{The temperature dependences of the zero-bias $R_{\Box }$ 
in the ''insulating'' regime ($V_{g}=-0.7V)$ at $B=0$ and $B=2T$. The solid
lines correspond to $R_{\square }(T)=33.3k\Omega \cdot \exp \left[ \left(
2.6K/T\right) ^{0.7}\right] $ at $B=0,$ and $R_{\square }(T)=25k\Omega \cdot
\exp \left[ \left( 1.57K/T\right) ^{0.7}\right] $ at $B=2T$. The inset shows 
$R=V/I$ at a fixed MC temperature $T=75mK$ and $B=2T$ versus the power $%
P=V\cdot I$ released in the device. The horizontal dashed line corresponds
to the zero-bias $R_{\square }$ at $T=82.5mK$.}
\label{Fig.2}
\end{center}
\end{figure}
\qquad 

With increase of the bias current, the {\it I-V} curves become nonlinear in
both WL and SL regimes. The insets in Figs. 1 and 2 show the resistance $R=V/I$ at a
fixed mixing chamber (MC) temperature $T$ versus the bias
current power, $P=V\cdot I$. It is instructive to compare for both regimes the power that
causes the decrease of $R$ that is equivalent to increase of $T$ by a fixed percentage 
(below we choose $10\%$). The
procedure is illustrated by the insets of Figs.1 and 2. The resistance,
measured at a fixed $T$ and different $I$, has been
compared with the zero-bias $R(1.1T)$. The corresponding values of $P$,
shown by the vertical arrows in the insets, are plotted for $T=0.05-1K$ in
Fig.3. The same measurements have been repeated in the SL regime for several
values of $B$. Within the experimental accuracy, no dependence $P(B)$ was
observed in the range $B=0-8T$ (the data for $B=2T$ are shown in Fig.3). Below we analyze
separately the data for the WL and SL regimes. 

\begin{figure}[th]
\begin{center}
\includegraphics[height=3.5in, width=3in]{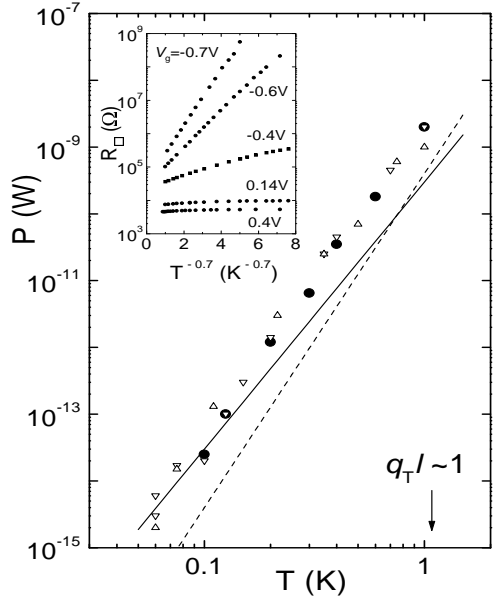}
\caption{The temperature dependence of the power $P=V\cdot I$, which causes
the increase of the electron temperature $T_{e}$ by 10\% over the MC
temperature $T$: $\bullet $ - in the WL\ regime ($V_{g}=0.4V$, $B=0$); $%
\bigtriangledown $ and $\triangle $ in the SL regime ($V_{g}=-0.7V$) at $B=0$
and $B=2T$, correspondingly. The dependences $P(T_{e}=1.1T)$ are calculated
for the electron cooling due to the piezoelectric coupling: the solid line -
for the disordered case (Eq. 2), the dashed line - for the clean case (Eq.
3). The inset: the zero-bias dependences $R_{\Box }(T)$ for several values 
of $V_{g}$.}
\label{Fig.3}
\end{center}
\end{figure}

{\bf The WL regime.}
The low-temperature nonlinear effects in the WL regime have been intensively studied in 
recent years \cite{hot-el,mittal}, they are accounted for by
the electron overheating. The hot-electron model
assumes that the non-equilibrium electron distribution function can be
characterized by an effective electron temperature $T_{e}$, the
electron-phonon interaction is the bottleneck in the energy transfer from
the electrons to the heat sink, and the phonons are in equilibrium with the
heat sink (the phonon temperature $T_{ph}$ is the same as $T$). One can find 
$T_{e}$ from comparison of $R(T)$ measured at different currents, provided
the zero-bias $R$ depends on $T_{e}$ only. 

All these
assumptions can be justified in our experiment. Indeed, at sub-Kelvin
temperatures, the electron-electron scattering rate is much greater than 
the electron-phonon one; this allows to introduce $T_{e}$. Both WL and interaction 
corrections are only $T_{e}$-dependend,
since the dominant phase-breaking mechanism is the quasi-elastic
electron-electron scattering \cite{gkrsw,prb-gersh,aak}. We have also verified that the
phonons in the GaAs chip remain in equilibrium with the mixing chamber (in
other words that the resistance drop at large $I$ is not due to heating of
the whole chip relative to the thermometer monitoring the MC temperature).
In this test, the zero-bias resistance of one of the devices (a phonon
''thermometer'') was measured at $T=0.1K$ as a function of the Joule heat
released in the other device on the same chip (a ''heater'') (see the inset
of Fig.4). The power required for a $10\%$-increase of the temperature of the
whole chip is by 3.5 orders of magnitude greater than the power that causes
nonlinear effects in the experiments when the same device combines the
functions of the ''heater'' and the ''thermometer''.

Since the outdiffusion of ''hot'' electrons in cooler leads can be neglected
for our samples at $T>0.1K$ \cite{prober,mittal}, the energy is transferred from
electrons to the heat sink due to the electron-phonon interaction only.
At $T\leqslant 1K$, the device enters the hydrodynamic regime $q_{t}l<1$ 
[$q_{t}=k_{B}T/\hbar u_{t}$ is the wave vector of a transverse phonon, $u_{t}$
is the transverse sound velocity ($\simeq 3\cdot 10^{3}m/s$ for GaAs), the
electron mean free path $l\sim 20nm$ at $V_{g}=0.4V$]. The low-$T$ data are
in agreement with calculations of the energy flow from the 2d electrons in
GaAs to the bulk phonons due to the piezoelectric coupling at $q_{t}l<1$ 
\cite{girvin}:

\begin{equation}
P[W]\simeq 7\cdot 10^{-2}\cdot \frac{e^{2}R_{\Box }}{h}\cdot A\cdot \left(
T_{e}^{4}-T^{4}\right) \left[ K^{4}\right] \text{ \ ,}  \label{girvin}
\end{equation}
where $R_{\Box }$ is the sheet resistance, and $A$ is the area of the device
(the solid line in Fig.3). At $T\sim 1K$, the experimental values of $P$
exceed by a factor of $4-5$ both (\ref{girvin}) and the result for the clean
case $q_{t}l>1$ \cite{price}:

\begin{equation}
P[W]\simeq 1.7\cdot 10^{6}\cdot n^{-1/2}\cdot A\cdot \left(
T_{e}^{5}-T^{5}\right) \left[ K^{5}\right] \text{ \ }  \label{price}
\end{equation}
(the dashed line in Fig.3). Our experimental dependence $P(T)$ is consistent
with the data for the GaAs heterostructures with larger $l\sim 0.3-1\mu m$,
obtained in Ref.\cite{mittal} at $T=0.1-0.4K$.

{\bf The SL regime.}
The striking similarity of the dependences $P(T)$ obtained in the
''metallic'' and ''insulating'' regimes indicate that the mechanism of 
nonlinearity is also the same on both sides of the WL-SL crossover.
Figure 4 shows that the hot-electron model describes very well the
dependences $R(P)$ in the SL regime \cite{1d}. It might be, of course, just
a coincidence: some field effects in the SL regime could produce, in
principle, similar nonlinearities. However, measurements in the magnetic
field rule out this possibility. Indeed, all the models of non-ohmic hopping
transport predict the dependence of the characteristic electric field $E$ on
the localization length $\xi $ \cite{shklovskii}. In the magnetic field,
which breaks the time-reversal symmetry within the localization domain, $\xi 
$ increases significantly \cite{efetov}: the exponentially strong negative
magnetoresistance, which has been observed for 1d and 2d $\delta $-doped
GaAs structures (see Fig. 2), is a signature of this $B$-induced growth of 
$\xi $ \cite{prl-gersh,prb-gersh,gkrsw}.  Despite of the increase of $\xi $
in strong magnetic fields, no dependence $P(B)$ was observed \cite
{3d}. 
\begin{figure}[th]
\begin{center}
\includegraphics[height=3.5in, width=3in]{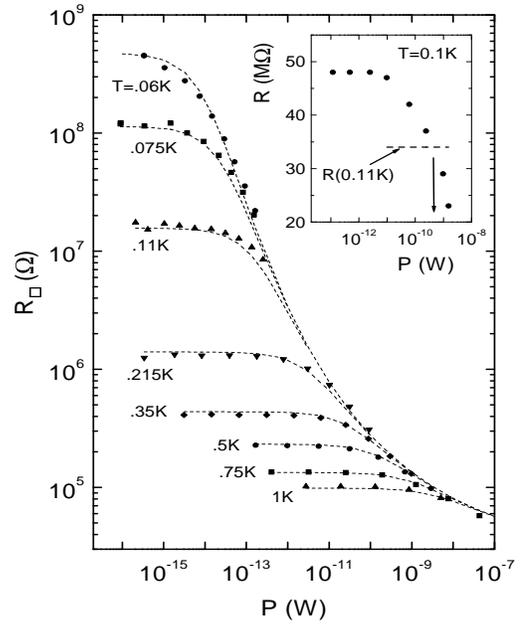}
\caption{The resistance $R=V/I$ in the SL regime ($V_{g}=-0.7V$, $B=2T$) at
different MC temperatures as a function of the power $P=V\cdot I$ released
in the device. The dashed lines are the fit $R_{\square }(T)=R_{\square }\exp %
\left[ \left( T_{o}/T_{e}\right) ^{0.7}\right] $ with $R_{\square
}=25k\Omega $, $T_{0}=1.57K$ (see Fig.2) and $T_{e}$ found from $P=3.7\cdot
10^{-9}\left[ W\right] \left( T_{e}^{4.5}-T^{4.5}\right) $, the best
approximation for the experimental data in Fig.3. The inset: 
the zero-bias $R$ in the SL regime ($V_{g}=-0.6V$, $T=0.1K$)
versus the power released in the other device on the same chip. The
horizontal dashed line corresponds to the zero-bias $R$ at $T=0.11K$.}
\label{Fig.4}
\end{center}
\end{figure}

The applicability of the hot-electron model to the SL regime indicates that:
a) the electron-phonon interaction remains the same on the ''insulating''
side of the crossover, and b) the resistance in the SL regime depends on the
electron temperature only. The former conclusion is not very surprising.
Indeed, the localization length, which can be estimated from the
magnetoresistance \cite{prl-gersh,prb-gersh,gkrsw}, is large close to the
crossover ($\xi \sim 0.15\mu m$ for $V_{g}=-0.7V$). The electron motion is
still diffusive at distances $\leq \xi $, with the $l$ values similar to that
in the WL regime. Since $q_{T} \xi \sim 1$ even at $T=50mK$,
one should not expect strong modification of the electron-phonon interaction
on the ''insulating'' side of the crossover. The latter conclusion, however,
implies that the hopping transport is not phonon-assisted, instead, it is
caused by electron-electron interactions.

Recently, the authors of Ref. \cite{khondaker} came to a similar conclusion
that hopping in {\it GaAs/AlGaAs}
heterostructures is not phonon-assisted. This conclusion was based 
on analysis of the prefactor $R_{0\Box }$ in Eq. (\ref{expfit}). 
Indeed, $R_{0\Box }$ is expected to depend on temperature and material properties
for the phonon-assisted hopping. Instead, the experimental $R_{0\Box }$
values are material-independent. They are usually close to the quantum resistance 
$R_{Q}=h/e^{2}$ \cite{dahm,khondaker,hsu,goldman}; this value is by several orders 
of magnitude smaller than the estimate for the phonon-assisted transport. In our experiment,
$R_{0\Box }$ is also close to $R_{Q}$ for both $B=0$ and strong magnetic fields
(see Fig. 2). 

Previously, the hot-electron model has been applied to the
non-ohmic effects in hopping conductivity of {\it three-dimensional}
heavily-doped {\it Ge} \cite{3d}. Our 2d
structures have two important features that help to attribute unambiguously the
observed nonlinear effects to the electron overheating: a) observation of
the WL-SL crossover allows direct comparison with the well-understood
''metallic'' regime, and b) for the 2d electron gas embedded in bulk {\it GaAs},
it is much easier to maintain equilibrium
between the phonons and the heat sink than in the case of the 3d uniformly-doped
samples.

To summarize, we have shown that the non-ohmic effects in the conductivity
of 2d Si$\ \delta $-doped GaAs structures, observed on both sides of the
WL-SL crossover, are caused by the electron overheating. The heat flow from
the hot 2d electrons to equilibrium 3d phonons in the WL and SL regimes is
well described by recent calculations for the piezoelectric coupling in the
hydrodynamic regime \cite{girvin}. The conductivity in the hopping regime
depends on the electron temperature rather than the phonon one, similar to
the WL regime. This observation provides strong evidence that electron
hopping in disordered systems with a large localization length is assisted
by the electron-electron interactions. 

We thank E. Abrahams, B. Altshuler, I. Aleiner, B. Shklovskii, M. Raikh, M. Reizer, 
and B. Spivak for
helpful disccusions. This work was partially supported by ARO-administered
MURI grant DAAD 19-99-1-0252 and by the Rutgers Office of the Research and
Sponsored Programs. D. R. gratefully acknowledges financial support of the 
DFG Graduiertenkolleg 384.


\begin{references}
\bibitem{saturation}  A plausible reason for absence of the $R(T)$
saturation in the SL regime is that the noise voltage $V_{n}$ applied to the
device is independent of the resistance for sufficiently large $R$ ($%
R>10^{4}\Omega $ even in the WL regime). Since the noise power $%
P=V_{n}^{2}/R $ required for electron overheating is the same in both WL and
SL regimes, the electron temperature is much less affected by the noise in
the SL regime, where $R$ is greater by $10^{5}$ than that in the WL regime.
\end{references}
\end{document}